\documentclass[showpacs,preprintnumbers,amsmath,amssymb]{revtex4}

\usepackage{graphicx}
\usepackage{dcolumn}
\usepackage{bm}

\newcommand{\bs}[1]{\mbox{\boldmath $#1$}}
\begin{document}


\title{
Monte Carlo Simulations in \\
Multibaric-Multithermal Ensemble 
}

\author{Hisashi Okumura$^{1,}$\footnote{Electronic address: hokumura@ims.ac.jp} 
and Yuko Okamoto$^{1,2,}$\footnote{Electronic address: okamotoy@ims.ac.jp}}
\address{
$^1\,$Department of Theoretical Studies \\
     Institute for Molecular Science \\ 
     Okazaki, Aichi 444-8585, Japan\\
$^2\,$Department of Functional Molecular Science \\
     The Graduate University for Advanced Studies \\ 
     Okazaki, Aichi 444-8585, Japan}



\begin{abstract}
We propose a new generalized-ensemble algorithm, which we refer to as 
the multibaric-multithermal Monte Carlo method. 
The multibaric-multithermal Monte Carlo simulations perform random walks widely 
both in volume space and in potential energy space. 
From only one simulation run, one can calculate
isobaric-isothermal-ensemble averages
at any pressure and any temperature.
We test the effectiveness of this algorithm by applying it to the 
Lennard-Jones 12-6 potential system 
with 500 particles.
It is found that a single simulation of the new method indeed gives
accurate average quantities in isobaric-isothermal ensemble for a wide
range of pressure and temperature.

\end{abstract}

\maketitle

%
Monte Carlo (MC) algorithm is one of the most widely used methods 
of computational physics. 
In order to realize desired statistical ensembles, 
corresponding MC techniques have been proposed
\cite{mrrtt53,mcd72,creu83,allen,frenkel}.
%
The first MC simulation was performed 
in the canonical ensemble 
in 1953 \cite{mrrtt53}. 
This method is called the Metropolis algorithm and widely used. 
The canonical probability distribution ${\rm P}_{NVT}(E)$ 
for potential energy $E$ is given by the product of 
the density of states $n(E)$ and 
the Boltzmann weight factor ${\rm e}^{-\beta_0 E}$: 
\begin{equation}
 {\rm P}_{NVT}(E) = n(E) {\rm e}^{-\beta_0 E}~, 
\label{eqn1}
\end{equation}
where $\beta_0$ is the inverse of the product of the Boltzmann
constant $k_B$ and temperature $T_0$ 
at which simulations are performed. 
Since $n(E)$ is a rapidly increasing function 
and the Boltzmann factor decreases exponentially, 
${\rm P}_{NVT}(E)$ is a bell-shaped distribution. 

The isobaric-isothermal MC simulation \cite{mcd72} 
is also extensively used. 
This is 
because most experiments are carried out 
under the constant pressure and constant temperature conditions. 
Both potential energy $E$ and volume $V$ fluctuate in this ensemble. 
The distribution ${\rm P}_{NPT}(E,V)$ 
for $E$ and $V$ is given by 
\begin{equation}
 {\rm P}_{NPT}(E,V) = n(E,V) {\rm e}^{-\beta_0 H}~. 
\label{eqn2}
\end{equation}
Here, the density of states $n(E,V)$ is given
as a function of both $E$ and $V$, 
and $H$ is the ``enthalpy'':
\begin{equation}
 H = E+P_0V~, 
\label{eqn3}
\end{equation}
where $P_0$ is the pressure at which simulations are performed. 
This ensemble has 
bell-shaped distributions in both $E$ and $V$. 

Besides the above physical ensembles, it is now almost a default to
simulate in artificial, generalized ensembles so that the
multiple-minima problem, or the broken ergodicity problem, in
complex systems can be overcome
(for a recent review, see Ref. \cite{mso01}).  
The multicanonical algorithm
\cite{bn91,bn92} 
is one of the most well known such methods in generalized ensemble. 
In multicanonical ensemble, a non-Boltzmann 
weight factor $W_{\rm mc}(E)$ is used. 
This multicanonical weight factor is characterized 
by a flat probability distribution ${\rm P_{mc}}(E)$: 
\begin{equation}
 {\rm P_{mc}}(E) = n(E) W_{\rm mc}(E) = {\rm constant}~, 
\label{eqn4}
\end{equation}
and thus a free
random walk is realized in the potential energy space. 
This enables the simulation to escape from any local-minimum-energy 
state and to sample the configurational space 
more widely than the conventional canonical MC algorithm. 
Another advantage is that one can obtain 
various canonical ensemble averages 
at any temperature from a single simulation run 
by the reweighting techniques \cite{fs88}. 
This method is now widely used in
complex systems such as proteins and glasses \cite{mso01}. 
However, it is difficult to compare the simulation conditions 
with experimental environments of constant pressure, 
since the simulations are performed in a fixed volume. 

In this Letter, we propose a new MC algorithm in which 
one can obtain various isobaric-isothermal ensembles 
from only one simulation. 
In other words, we introduce the idea of the multicanonical technique 
into the isobaric-isothermal ensemble MC method. 
We call this method the multibaric-multithermal algorithm. 
This MC simulation performs random walks 
in volume space as well as in potential energy space. 
As a result, this method has the following advantages: 
(1) It allows the simulation to escape from any local-minimum-energy 
state and to sample the configurational space 
more widely than the conventional isobaric-isothermal method. 
(2) One can obtain various isobaric-isothermal ensembles 
not only at any temperature, as in the multicanonical algorithm, 
but also at any pressure from only one simulation run. 
(3) One can control pressures and temperatures 
similarly to real experimental conditions. 


%
In the multibaric-multithermal ensemble, 
every state is sampled by a weight factor 
$W_{\rm mbt}(E,V)\equiv \exp \{-\beta_0 H_{\rm mbt}(E,V)\}$ 
($H_{\rm mbt}$ is referred to as the multibaric-multithermal enthalpy) 
so that a uniform distribution in both potential energy space 
and volume space is obtained: 
\begin{equation}
 {\rm P_{mbt}}(E,V) = n(E,V) W_{\rm mbt}(E,V) = {\rm constant}~. 
\label{eqn5}
\end{equation}
We call $W_{\rm mbt}(E,V)$ the multibaric-multithermal weight factor. 

In order to perform the multibaric-multithermal MC simulation, 
we follow 
the conventional isobaric-isothermal MC techniques \cite{mcd72}. 
In this method, we perform Metropolis sampling on the scaled coordinates 
${\bs s}_i = L^{-1} {\bs r}_i$ 
(${\bs r}_i$ are the real coordinates) 
and the volume $V$ (here, the particles are placed in a cubic box of
a side of size $L \equiv \sqrt[3]{V}$). 
The trial moves of the scaled coordinates 
from ${\bs s}_i$ to ${\bs s'}_i$ and 
of the volume from $V$ to $V'$ are generated by uniform random numbers. 
The enthalpy is accordingly changed 
from $H(E({\bs s}^{(N)},V),V)$ 
to $H'(E({\bs s}'^{(N)},V'),V')$ 
by these trial moves. 
The trial moves will be accepted with the probability 
\begin{equation}
 {\rm acc(o \rightarrow n) = min} 
 (1,\exp[-\beta_0 \{ 
 H' - H - N k_B T_0 \ln(V'/V)
 \}])~, 
 \label{accibt:eq}
\end{equation}
where $N$ is the total number of particles in the system.

Replacing $H$ by $H_{\rm mbt}$, 
we can perform the multibaric-multithermal MC simulation. 
The trial moves of ${\bs s}_i$ and $V$ are generated in the same way 
as in the isobaric-isothermal MC simulation. 
The multibaric-multithermal enthalpy is changed 
from $H_{\rm mbt}(E({\bs s}^{(N)},V),V)$ 
to $H'_{\rm mbt}(E({\bs s}'^{(N)},V'),V')$ 
by these trial moves. 
The trial moves will now be accepted with the probability 
\begin{equation}
 {\rm acc(o \rightarrow n) = min} 
 (1,\exp[-\beta_0 \{ 
 H'_{\rm mbt} - H_{\rm mbt} - N k_B T_0 \ln(V'/V)
 \}])~. 
 \label{acc:eq}
\end{equation}
The multibaric-multithermal 
probability distribution ${\rm P_{mbt}}(E,V)$ 
is obtained by this scheme.

In order to calculate the isobaric-isothermal ensemble average, 
we employ the reweighting techniques \cite{fs88}. 
The probability distribution ${\rm P}_{NPT}(E,V;T,P)$ 
at any temperature $T$ and any pressure $P$ 
in the isobaric-isothermal ensemble is given by
\begin{equation}
 {\rm P}_{NPT}(E,V;T,P) = 
 \frac
 {{\rm P_{mbt}}(E,V) \ W^{-1}_{\rm mbt}(E,V) \ 
  {\rm e}^{-\beta (E+PV)}}
 {\displaystyle{\int dV \ \int dE \ 
  {\rm P_{mbt}}(E,V) \ W^{-1}_{\rm mbt}(E,V) \ 
  {\rm e}^{-\beta (E+PV)}}}~. \label{mbtrwt:eq}
\end{equation}
The expectation value of a physical quantity $A$ 
at $T$ and $P$ is estimated from 
\begin{eqnarray}
 <A>_{NPT} 
 &=& \int dV \ \int dE \ A(E,V) \ {\rm P}_{NPT} (E,V;T,P)~, \nonumber \\
 &=& \frac
 {<A(E,V) W^{-1}_{\rm mbt}(E,V) 
 {\rm e}^{-\beta (E+PV)}>_{\rm mbt}}
 {<W^{-1}_{\rm mbt}(E,V) 
 {\rm e}^{-\beta (E+PV)}>_{\rm mbt}}~, 
 \label{mbtave:eq}
\end{eqnarray}
where $<\cdots>_{\rm mbt}$ is 
the multibaric-multithermal ensemble average. 

After having given the formalism of the multibaric-multithermal 
algorithm, let us now describe the process 
for determining the weight factor $W_{\rm mbt}(E,V)$. 
This is obtained by the usual iteration of short simulations
\cite{bc92,lee93,oh95}. 
The first simulation is carried out at $T_0$ and $P_0$ 
in the isobaric-isothermal ensemble. 
Namely, we use 
\begin{equation}
 W^{(1)}_{\rm mbt}(E,V) = \exp \{-\beta_0 H^{(1)}_{\rm mbt}(E,V)\}~,
\label{eqn10}
\end{equation}
where
\begin{equation}
 H^{(1)}_{\rm mbt}(E,V) = E + P_0 V~.
\label{eqn11}
\end{equation}
The $i$-th simulation is performed with 
the weight factor $W^{(i)}_{\rm mbt}(E,V)$ and let
${\rm P}^{(i)}_{\rm mbt}(E,V)$ be the obtained
distribution.
The $(i+1)$-th weight factor $W^{(i+1)}_{\rm mbt}(E,V)$ is then given by
\begin{equation}
 W^{(i+1)}_{\rm mbt}(E,V) = 
 \exp \{-\beta_0 H^{(i+1)}_{\rm mbt}(E,V)\}~, \label{wimbt:eq}
\end{equation}
where
\begin{equation}
 H^{(i+1)}_{\rm mbt}(E,V) 
 =
 \left\{
 \begin{array}{ll}
   H^{(i  )}_{\rm mbt}(E,V) + 
   k_B T_0 \ln {\rm P}^{(i)}_{\rm mbt}(E,V)~, 
   &
   \hspace{-8mm}\qquad {\rm for}~~ {\rm P}^{(i)}_{\rm mbt}(E,V) > 0~, \\
   \noalign{\vskip0.2cm}
   H^{(i  )}_{\rm mbt}(E,V)~, 
   &
   \hspace{-8mm}\qquad {\rm for}~~ {\rm P}^{(i)}_{\rm mbt}(E,V) = 0~. 
   \label{himbt:eq} 
 \end{array} 
 \right. \\
\end{equation}
For convenience, we make $E$ and $V$ discrete into bins 
and use histograms to calculate ${\rm P}^{(i)}_{\rm mbt}(E,V)$. 
We iterate this process until a reasonably flat 
distribution ${\rm P}^{(i)}_{\rm mbt}(E,V)$ 
is obtained.
After an optimal weight factor is determined, a 
long simulation is performed to sample the configurational space. 


%
We now present the results of our multibaric-multithermal
simulation.
We considered a Lennard-Jones 12-6 potential system. 
We used 500 particles ($N=500$)
in a cubic unit cell with periodic boundary conditions. 
The length and the energy are scaled in units of 
the Lennard-Jones diameter $\sigma$ and 
the minimum value of the potential $\epsilon$, respectively. 
We use an asterisk ($*$) for the reduced quantities 
such as 
the reduced length $r^* = r/\sigma$, 
the reduced temperature $T^*=k_{\rm B}T/\epsilon$, 
the reduced pressure $P^*=P\sigma^3/\epsilon$, and 
the reduced number density $\rho^*=\rho \sigma^3$ ($\rho \equiv N/V$). 

We started the multibaric-multithermal weight factor determination
of Eqs. (\ref{wimbt:eq}) and (\ref{himbt:eq}) from a regular isobaric-isothermal
simulation
at $T_0^* = 2.0$ and $P_0^* = 3.0$. 
These temperature and pressure are respectively higher than 
the critical temperature $T_{\rm c}^*$ and 
the critical pressure $P_{\rm c}^*$ 
\cite{oku00jcp, oku01jpsj4, 
pan00, cai98, ppa98, oku01jpsj7}. 
Recent reliable data are
$T_{\rm c}^*=1.3207(4)$ and $P_{\rm c}^*=0.1288(5)$ \cite{oku01jpsj7}. 
The cutoff radius $r^*_{\rm c}$ was taken to be $L^*/2$. 
A cut-off correction was added for the pressure and
potential energy. 
In one MC sweep we made the trial moves of all particle coordinates and the volume 
($N+1$ trial moves altogether).  For each trial move the Metropolis evaluation
of Eq. (\ref{acc:eq}) was made.
In order to obtain a flat probability 
distribution ${\rm P}_{\rm mbt}(E,V)$, 
we carried out the MC simulations 
of 100,000 MC sweeps and 
iterated the process of Eqs. (\ref{wimbt:eq}) and (\ref{himbt:eq}). 
In the present case, it was required to make 12 iterations to
get an optimal weight factor
$W_{\rm mbt}(E,V)$.
We then performed a long multibaric-multithermal MC simulaton 
of 400,000 MC sweeps 
with this $W_{\rm mbt}(E,V)$. 

For the purpose of comparisons of the new method with the conventional one, 
we also performed the conventional isobaric-isothermal MC simulations 
of 100,000 MC sweeps
with 500 Lennard-Jones 12-6 potential particles 
at several sets of temperature and pressure. 
The temperature ranged from $T^*=1.6$ to 2.6 and 
the pressure from $P^*=2.2$ to 3.8. 

In order to estimate the statistical accuracies, 
we performed these MC simulations 
from four different initial conditions 
in both multibaric-multithermal and 
isobaric-isothermal simulations. 
The error bars were calculated by the standard deviations 
from these different simulations. 

\begin{figure}[h]
\includegraphics[width=15cm,keepaspectratio]{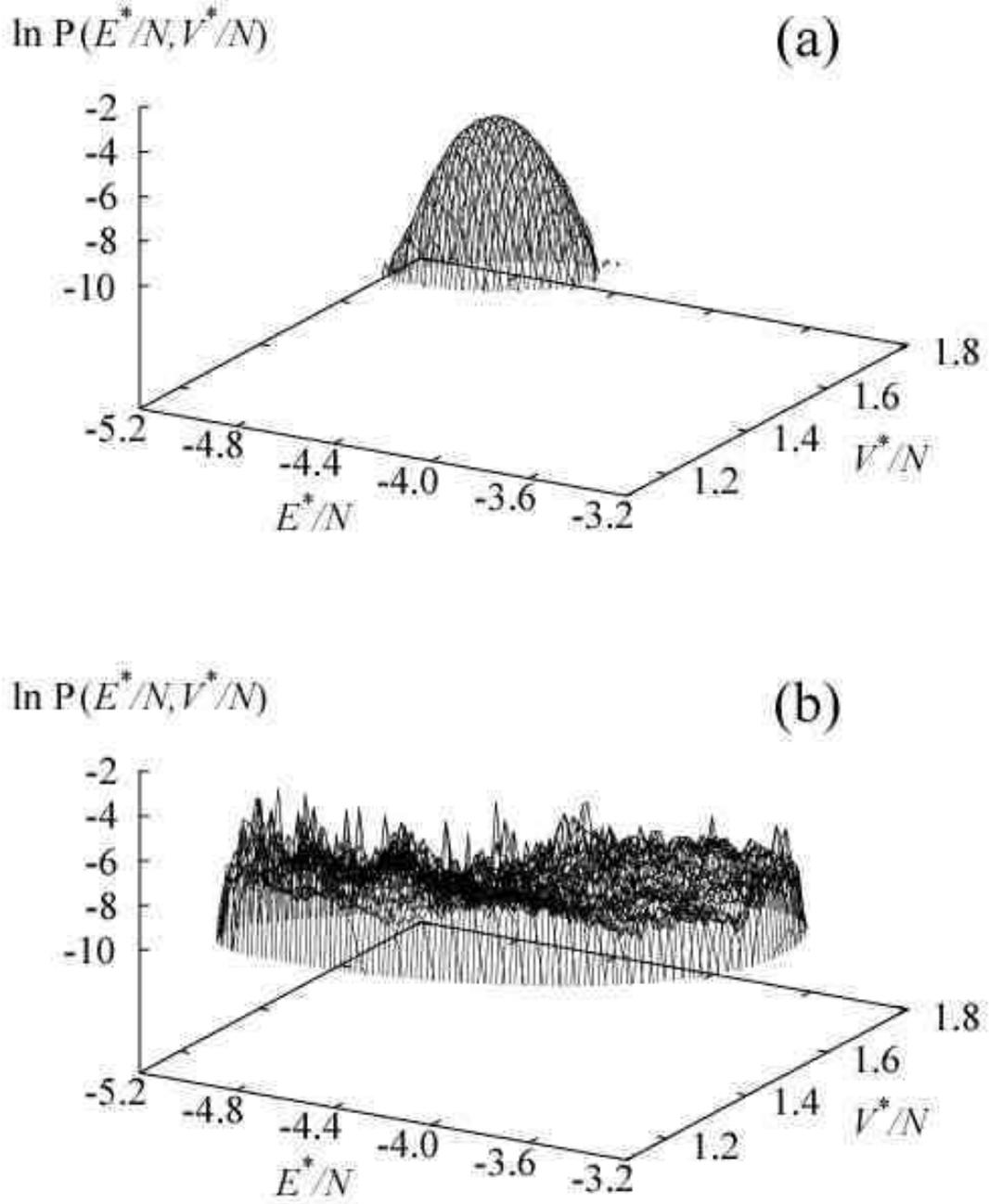}
\caption{
(a) The probability distribution ${\rm P}_{NPT}(E^*/N,V^*/N)$ 
in the isobaric-isothermal simulation 
at $(T^*,P^*)=(T_0^*,P_0^*)=(2.0,3.0)$ and
(b) the probability distribution ${\rm P_{\rm mbt}}(E^*/N,V^*/N)$ 
in the multibaric-multithermal simulation. 
}
\label{dis:fig}
\end{figure}

%
Figure \ref{dis:fig} shows the probability distributions of 
$E^*/N$ and $V^*/N$. 
Figure \ref{dis:fig}(a) is the probability 
distribution ${\rm P}_{NPT}(E^*/N,V^*/N)$ 
from the isobaric-isothermal simulation first carried out in the
process of Eqs. (\ref{eqn10}) and (\ref{eqn11})
(i.e., $T_0^*=2.0$ and $P_0^*=3.0$). 
It is a bell-shaped distribution. 
On the other hand, Fig. \ref{dis:fig}(b) is the probability 
distribution ${\rm P_{\rm mbt}}(E^*/N,V^*/N)$ 
from the multibaric-multithermal simulation finally performed. 
It shows a flat distribution, and  
the multibaric-multithermal MC simulation indeed 
sampled the configurational space 
in wider ranges of energy and volume 
than the conventional isobaric-isothermal MC simulation. 

\begin{figure}[h]
\includegraphics[width=17cm,keepaspectratio]{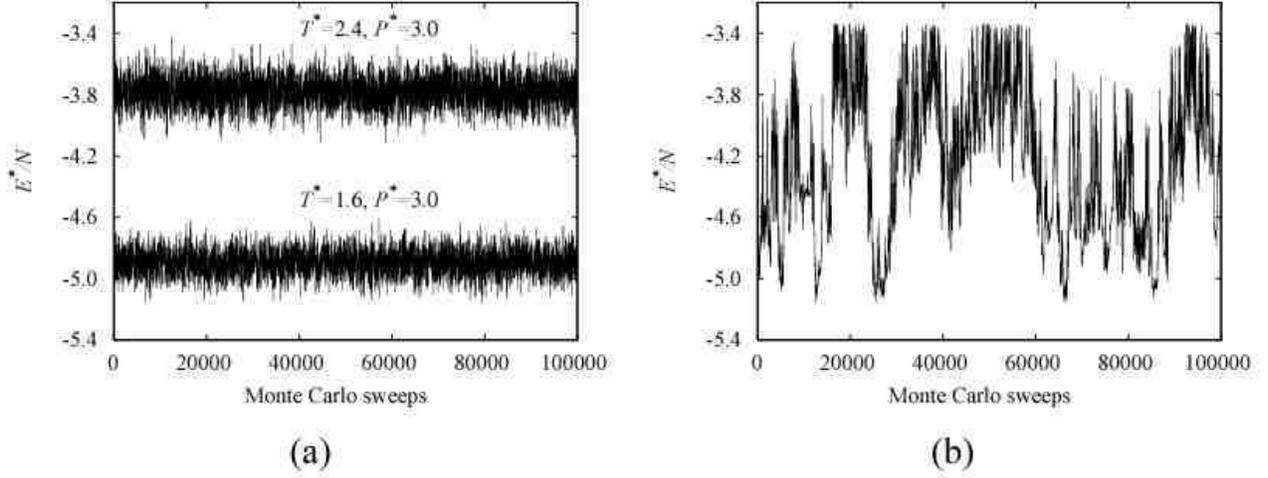}
\caption{
The time series of $E^*/N$ from 
(a) the conventional isobaric-isothermal MC simulations 
at $(T^*,P^*)=(2.4,3.0)$ and at $(T^*,P^*)=(1.6,3.0)$ and 
(b) the multibaric-multithermal MC simulation. 
}
\label{ene:fig}
\end{figure}

Figure \ref{ene:fig} shows the time series of $E^*/N$. 
Figure \ref{ene:fig}(a) gives the results of the conventional
isobaric-isothermal simulations at
$(T^*,P^*)=(1.6,3.0)$ and (2.4,3.0), while
Figure \ref{ene:fig}(b) presents those of the multibaric-multithermal
simulation.
The potential energy fluctuates in narrow ranges 
in the conventional isobaric-isothermal MC simulations. 
They fluctuate only 
in the ranges of $E^*/N = -4.0 \sim -3.6$ and $E^*/N = -5.1 \sim -4.7$ 
at the higher temperature of $T^*=2.4$ and 
at the lower temperature of $T^*=1.6$, respectively. 
On the other hand, 
the multibaric-multithermal MC simulation performs a random walk 
that covers a wide energy range.

\begin{figure}[h]
\includegraphics[width=17cm,keepaspectratio]{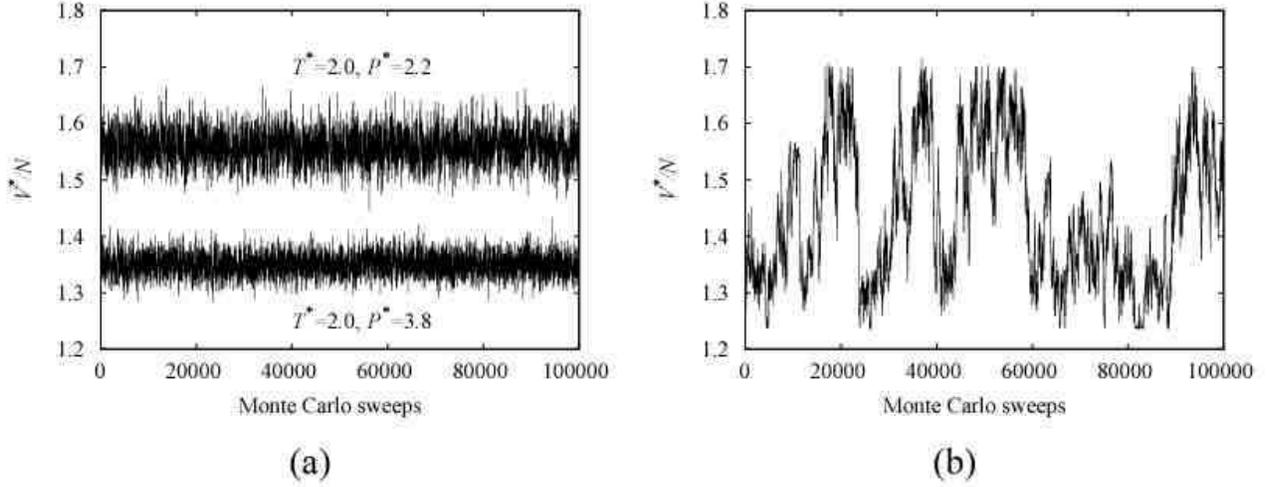}
\caption{
The time series of $V^*/N$ from 
(a) the conventional isobaric-isothermal MC simulations 
at $(T^*,P^*)=(2.0,2.2)$ and at $(T^*,P^*)=(2.0,3.8)$ and 
(b) the multibaric-multithermal MC simulation. 
}
\label{vol:fig}
\end{figure}

A similar situation is observed in 
the time series of $V^*/N$. 
In Fig. \ref{vol:fig}(a) we show the results of the conventional
isobaric-isothermal simulations at
$(T^*,P^*)=(2.0,2.2)$ and (2.0,3.8), while in
Figure \ref{vol:fig}(b) we give those of the multibaric-multithermal
simulation.

The volume fluctuations 
in the conventional isobaric-isothermal MC simulations 
are only in the range of 
$V^*/N = 1.3 \sim 1.4$ and $V^*/N = 1.5 \sim 1.6$ 
at $P^*=3.8$ and at $P^*=2.2$, respectively. 
On the other hand, 
the multibaric-multithermal MC simulation performs a random walk 
that covers even a wider volume range.

\begin{figure}[h]
\includegraphics[width=15cm,keepaspectratio]{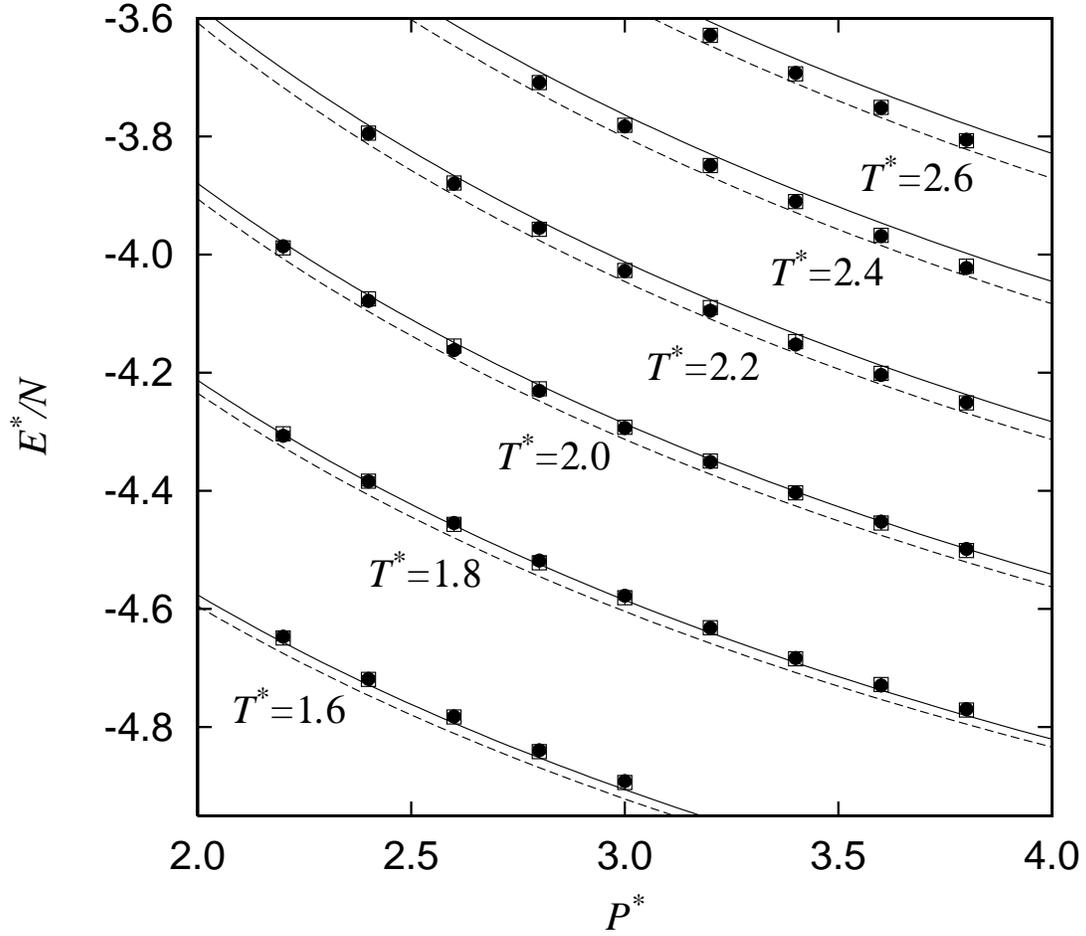}
\caption{
Average potential energy per particle $<E^*/N>_{NPT}$ at various 
temperature and pressure values. 
Filled circles: Multibaric-multithermal MC simulations. 
Open squares: Conventional isobaric-isothermal MC simulations. 
Solid line: Equation of states calculated by Johnson et al. [19]. 
Broken line: Equation of states calculated by Sun and Teja [20]. 
}
\label{rwte:fig}
\end{figure}

\begin{figure}[h]
\includegraphics[width=15cm,keepaspectratio]{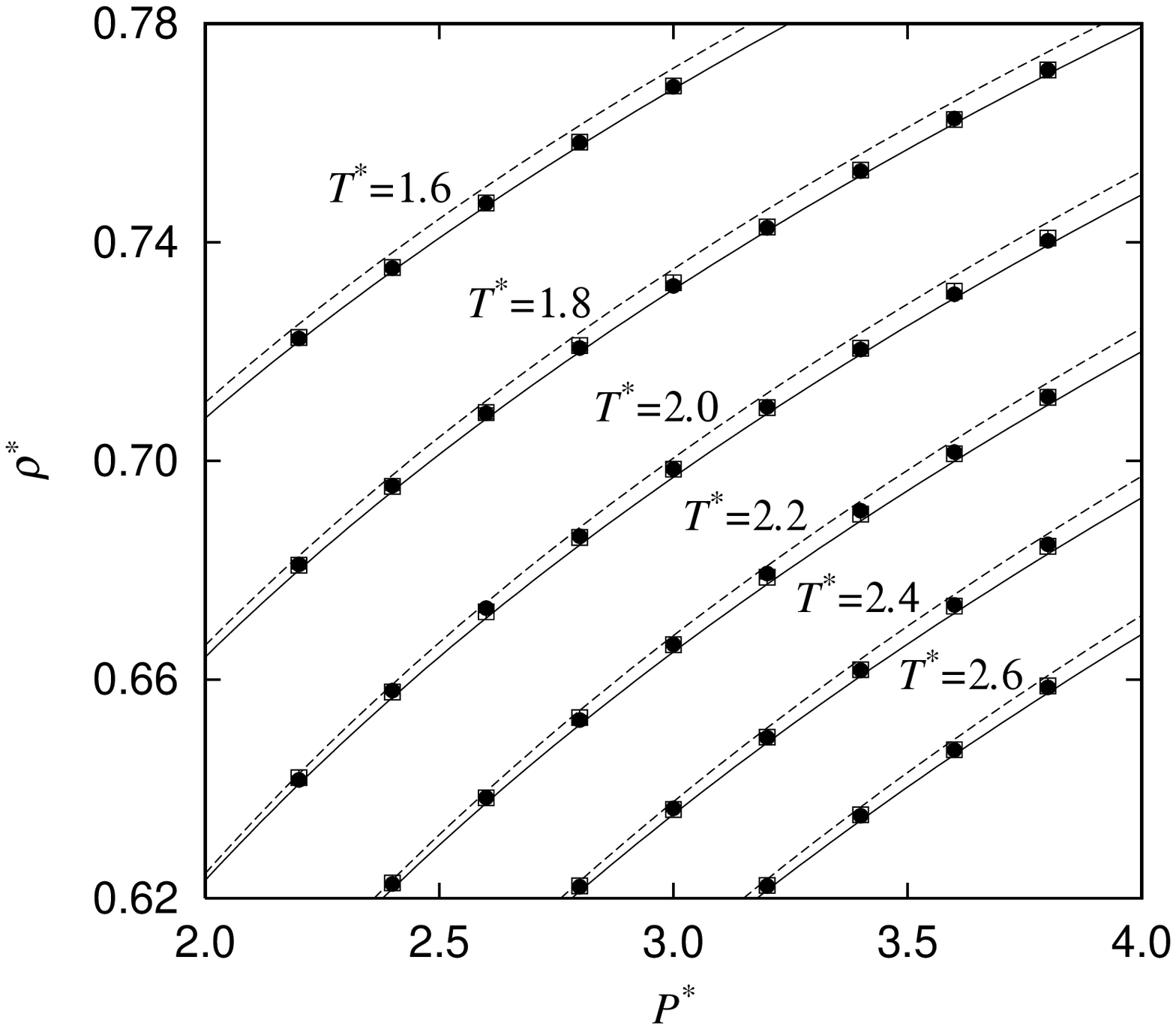}
\caption{
Average density $<\rho^*>_{NPT}$ at various temperature and pressure values. 
See the caption of Fig. 4 for details. 
}
\label{rwtd:fig}
\end{figure}

We calculated the ensemble averages of potential energy per particle,
$<E^*/N>_{NPT}$, 
and density, $<\rho^*>_{NPT}$, at various temperature and pressure values
by the reweighting techniques of Eq. (\ref{mbtave:eq}).
They are shown in Fig. \ref{rwte:fig} and in Fig. \ref{rwtd:fig}, respectively. 
The error bars are smaller than the plots for both cases. 
The agreement between the multibaric-multithermal data and 
isobaric-isothermal data are excellent 
in both $<E^*/N>_{NPT}$ and $<\rho^*>_{NPT}$. 

The important point is that 
we can obtain any desired isobaric-isothermal distribution 
in wide temperature and 
pressure ranges ($T^*=1.6 \sim 2.6$, $P^*=2.2 \sim 3.8$) 
from a single simulation run by the multibaric-multithermal MC algorithm. 
This is an outstanding advantage over
the conventional isobaric-isothermal MC algorithm, 
in which simulations have to be carried out 
separately at each temperature and pressure, because the reweighting
techniques based on the isobaric-isothermal simulations
can give correct results only for
narrow ranges of temperature and pressure values.


Figures \ref{rwte:fig} and \ref{rwtd:fig} also show 
two equations of states of the Lennard-Jones 12-6 potential fluid. 
One is determined by Johnson et al. \cite{jzg93} and 
the other by Sun and Teja \cite{sj93}. 
These equations are determined by 
fitting procedure to the molecular simulation results. 
Our multibaric-multithermal simulation results agree very well
with those of these equations. 
Investigating in more detail, however, 
the two equations give slightly different results.
Most of our data lie in between them. 

%
In conclusion, we proposed a new MC algorithm that is based
on multibaric-multithermal ensemble. 
We applied this method to the Lennard-Jones 12-6 potential system. 
The advantage of this method is that 
the simulation performs random walks 
in both potential energy space and volume space and 
sample the configurational space much more widely 
than the conventional isobaric-isothermal MC method. 
Therefore, one can obtain various isobaric-isothermal 
ensemble averages at any desired temperature and pressure from 
only one simulation run. 
This allows one to
specify a pressure and to 
compare simulation conditions directly with those of real experiments. 
The multibaric-multithermal algorithm will thus be of great use for 
investigating a large variety of complex systems 
such as proteins, polymers, supercooled liquids, and glasses. 
It will be particularly useful for the study of,
for example, pressure induced phase transitions. 

%
We would like to thank M. Mikami of National
Institute of Advanced Industrial Science and Technology
and U.H.E. Hansmann of Michigan
Technological University for useful discussions at the
early stage of the present work.

%
%
%

\clearpage


%
%
%
%
%
%
%
%
%
%

\begin{thebibliography}{99}
%
\bibitem{mrrtt53}
N. Metropolis, A. W. Rosenbluth, M. N. Rosenbluth, A. H. Teller, and E. Teller,  
{\it J. Chem. Phys.} {\bf 21}, 1087 (1953).
%
\bibitem{mcd72}
I. R. McDonald, 
{\it Mol. Phys.} {\bf 23}, 41 (1972).
%
\bibitem{creu83}
M. Creutz, 
{\it Phys. Rev. Lett.} {\bf 50}, 1411 (1983).
%
\bibitem{allen}
M. P. Allen and D. J. Tildesley,
{\it Computer Simulation of Liquids},
(Clarendon Press, Oxford, 1987) p. 110.
%
\bibitem{frenkel}
D. Frenkel and B. Smit,
{\it Understanding Molecular Simulation From Algorithms to Applications},
(Academic Press, San Diego, 2002) p. 111.
%
\bibitem{mso01}
A. Mitsutake, Y. Sugita, and Y. Okamoto,
{\it Biopolymers (Peptide Science)} {\bf 60}, 96 (2001).
%
\bibitem{bn91}
B. A. Berg and T. Neuhaus,
{\it Phys. Lett.} {\bf B267}, 249 (1991).
%
\bibitem{bn92}
B. A. Berg and T. Neuhaus,
{\it Phys. Rev. Lett.} {\bf 68}, 9 (1992).
%
%
%
%
%
\bibitem{fs88}
A. M. Ferrenberg and R. H. Swendsen,
{\it Phys. Rev. Lett.} {\bf 61}, 2635 (1988); 
$ibid$. {\bf 63}, 1658(E) (1989). 
%
\bibitem{bc92}
B. A. Berg and T. Celik,
{\it Phys. Rev. Lett.} {\bf 69}, 2292 (1992).
%
\bibitem{lee93}
J. Lee,
{\it Phys. Rev. Lett.} {\bf 71}, 211 (1993); {\it ibid.} 2353(E) (1993).
%
\bibitem{oh95}
Y. Okamoto and U. H. E. Hansmann, 
{\it J. Phys. Chem.} {\bf 99}, 11276 (1995).
%
%
\bibitem{oku00jcp}
H. Okumura and F. Yonezawa,
{\it J. Chem. Phys.} {\bf 113}, 9162 (2000).
%
%
\bibitem{oku01jpsj4}
H. Okumura and F. Yonezawa,
{\it J. Phys. Soc. Jpn.} {\bf 70}, 1006 (2001).
%
\bibitem{pan00}
A. Z. Panagiotopoulos,
{\it J. Phys.: Condens. Matter} {\bf 12}, R25 (2000).
%
%
%
%
\bibitem{cai98}
J. M. Caillol, 
{\it J. Chem. Phys.} {\bf 109}, 4885 (1998).
%
\bibitem{ppa98}
J. J. Potoff and A. Z. Panagiotopoulos,
{\it J. Chem. Phys.} {\bf 109}, 10914 (1998).
%
\bibitem{oku01jpsj7}
H. Okumura and F. Yonezawa,
{\it J. Phys. Soc. Jpn.} {\bf 70}, 1990 (2001).
%
\bibitem{jzg93}
J. K. Johnson, J. A. Zollweg, and K. E. Gubbins,
{\it Mol. Phys.} {\bf 78}, 591 (1993). 
%
\bibitem{sj93}
T. Sun and A. S. Teja,
{\it J. Phys. Chem.} {\bf 100}, 17365 (1996). 
%
\end{thebibliography}
\end{document}